\documentclass[aps,
english,preprintnumbers,nofootinbib,
twocolumn]{revtex4-1}

\usepackage{amsfonts,amsmath,amssymb}
\usepackage{graphicx}
\usepackage[utf8]{inputenc}
\usepackage{hyperref}
\usepackage{babel}

\usepackage{xcolor}

\newcommand\be{\begin{equation}}
\newcommand\ee{\end{equation}}
\newcommand\bea{\begin{eqnarray}}
\newcommand\eea{\end{eqnarray}}

\begin{document}

\title{A universal approach to Krylov State and Operator complexities}

\author{Mohsen Alishahiha${}^a$\email{alishah@ipm.ir} and, Souvik Banerjee${}^{b,c}$
\email{souvik.banerjee@physik.uni-wuerzburg.de}}
\affiliation{${}^a$ School of Physics, Institute for Research in Fundamental Sciences (IPM),\\
	P.O. Box 19395-5531, Tehran, Iran\\ ${}^b$ Institut f{\"u}r Theoretische Physik und Astrophysik,
	Julius-Maximilians-Universit{\"a}t W{\"u}rzburg,\\ Am Hubland, 97074 W{\"u}rzburg, Germany\\ ${}^c$ W\"urzburg-Dresden Cluster of Excellence ct.qmat\\}

\begin{abstract}
We present a general framework in which both Krylov state and operator complexities
can be put on the same footing.  In our formalism, the Krylov complexity is defined 
in terms of the density matrix of the associated state which, for the operator complexity, 
lives on a doubled Hilbert space obtained through the channel-state map. 
This unified definition of complexity in terms of the density matrices enables us to 
extend the notion of Krylov complexity, to subregion or mixed state complexities and also naturally 
to the Krylov mutual complexity. We show that this framework also encompasses nicely the holographic notions 
of complexity. 
\end{abstract}

\maketitle

\section{Introduction}
Quantum mechanics provides us with two apparently different yet natural ways 
to understand the complexity of a system through time-evolution.
In the Schr\"odinger picture which allows the state to be time-dependent, 
complexity measures the mixing of the initial state with other states through 
the time-evolution, which for a time independent Hamiltonian $H$ takes the 
form
\be\label{St}
|\psi(t)\rangle=e^{iHt}|\psi(0)\rangle = \sum_{n=0}^{\infty} 
\frac{(i t)^n}{n!} |\psi^{(n)}\rangle,
\ee
where $|\psi^{(n)}\rangle \equiv H^n |\psi(0)\rangle$. Thus, in this 
notation, complexity of the final state amounts to understanding the spread 
of the corresponding wavefunction in a fixed basis. The choice of an optimal 
basis here is a bit tricky. One interesting way proposed in 
\cite{Balasubramanian:2022tpr} where the minimization of the spreading of the 
wavefunction determines the optimal basis. This notion of state complexity 
was coined as the ``spread complexity''.

Alternatively, one may wish to switch to the Heisenberg picture where the 
time evolution is attributed to the operator instead. This is realised by 
noting the expectation value of an operator in the complexified state
\be
\langle \psi(t)| {\cal O} |\psi(t)\rangle=\langle \psi(0)| e^{-iHt}{\cal O} 
e^{iHt} |\psi(0)\rangle
\ee
and immediately recognizing the time dependent operator
\be
{\cal O}(t)=e^{-iHt}{\cal O} e^{iHt}\, = \sum_{n=0}^{\infty} 
\frac{(i t)^n}{n!} {\cal O}^{(n)},
\ee
where ${\cal O}^{(n)}$ are understood as nested structures of operators, 
${\cal O}^{(n)}\equiv [H,\cdots,[H,{\cal O}]\cdots]$ and provide the notion 
of operator complexity by interpreting ${\cal O}(t)$ as the operator wave 
function evolved by a Liouvillian superoperator ${\cal L}$ 
\be
{\cal O}(t) = e^{i {\cal L} t} {\cal O},
\ee
with ${\cal L} = [H,\cdot]$. In this notation, ${\cal O}^{(n)} \equiv 
{\cal L}^n {\cal O}$ which determines the mixing of operators. In order to 
compute complexity corresponding to the growth of the operator, one uses the 
Lanczos algorithm \cite{Lanczos:1950zz} to construct an optimal basis 
\cite{viswanath1994recursion}, known in literature as the Krylov basis. The 
corresponding operator complexity is termed as the Krylov complexity 
\cite{Parker:2018yvk}.

Last but not the least, in the present decade there has been one more entry 
in the world of complexity in form of the holographic complexity. As its name 
suggests, this notion of complexity arose out of the curiosity to understand 
the interior of a black hole spacetime in the light of AdS/CFT correspondence 
\cite{Maldacena:1997re, Witten:1998qj,Gubser:1998bc}. In particular, the fact 
that the volume the interior of the black hole keeps growing even after 
attaining thermal equilibrium \cite{Susskind:2014rva,Susskind:2018pmk} is 
very much reminiscent of the nature of complexity of a finite entropic 
fast-scrambling system.

Motivated by this striking similarity, a holographic definition of complexity 
was proposed as the volume of the maximal slice in the interior of the black 
hole. This proposal is celebrated in the name ``Complexity = Volume" (CV) 
conjecture \cite{Susskind:2014rva, Stanford:2014jda}. An efficient formalism 
to study this interior volume in two dimensional theory of gravity was 
developed in \cite{Iliesiu:2021ari, Alishahiha:2022kzc, Alishahiha:2022exn} 
which produces the expected behaviour of the late-time linear growth and 
eventual saturation of complexity. 
 
 This observation was formalized in a more general context in 
 \cite{Alishahiha:2022nhe}, which, based on the Eigenstate Thermalization 
 Hypothesis (ETH) \cite{Deutsch, Srednicki}, classified the possible 
 candidates for complexity which exhibit a linear growth at late time. It was 
 also shown that this class of observables naturally includes the expectation 
 value of the quenched length operator defined 
 in \cite{Iliesiu:2021ari, Alishahiha:2022kzc, Alishahiha:2022exn}.

It is a daunting task to bring all these apparently different notions of 
complexity under the same umbrella. And this is precisely the motivation of 
the present work. 

In the process of doing so, we will first present, {\bf in section 
\ref{sec2}},  a general framework to study complexity of a given state. This 
state can either be a given autonomous state or could as well be a state 
evolved under the evolution by a general Hermitian operator which is not 
necessarily the Hamiltonian of the system. In either case, for a given 
orthonormal and ordered basis, we can define a label operator and 
subsequently a number obtained by tracing the former over the density matrix 
corresponding to the given or the evolved state. This number is minimized 
over the space of operators used to generate sets of the orthonormal basis 
through Gram-Schmidt process, to define the complexity of the state.

These operators can be identified as different quantum gates in accordance 
with the quantum informatic definition of circuit complexity \cite{Nielsen}. 
When the state is created through a unitary evolution, this definition of 
complexity naturally yields the Krylov complexity. 

We show that, for the Hamiltonian evolution, the class of observables 
identified in \cite{Alishahiha:2022nhe}, with the desired pole structure to 
produce the late-time linear growth of complexity, can be identified, 
naturally, with the expectation value of the label operator mentioned above. 
We further show that our formalism also naturally provides the late time 
saturation of complexity following the linear growth. 

{\bf In section \ref{sec3}}, we develop the formalism to study operator 
complexity. The structure of the label operator dictates a mapping of the 
space of operators to a doubled Hilbert space endowed with a inner product 
structure. This enables us to recast the operator complexity associated with 
any given operator to the complexity of a state in the doubled Hilbert space.
This definition is also consistent with the Liouvillian evolution for 
operator complexity.

From the perspective of quantum information, the mapping to the doubled 
Hilbert space is a realisation of the channel-state map. However, we 
demonstrate that it has a beautiful interpretation in axiomatic quantum field 
theory and reveals a deeper structure of entanglement in the Hilbert space 
which turns out to be pivotal in understanding the holographic notion of 
complexity.  

{\bf In section \ref{sec4}} we discuss a novel biproduct of our generalized 
formalism for studying the state and the operator complexities. Since we 
defined complexity in terms of density matrices, it fits in as the ideal 
candidate to study subregion complexity using reduced density matrix of a 
given subregion. Definition of subregion Krylov complexity reveals some more 
interesting and fundamental aspects connecting quantum entanglement and the 
growth of complexity.

Finally, we conclude {\bf in section \ref{sec5}} with some interesting open 
questions and some works in progress.

\section{Complexity : The general framework}
\label{sec2}
Be it for a state or for an operator, the general strategy to compute 
complexity comprises of the following steps - i) start with an initial state 
(operator), ii)
allow it to spread  over some state (operator) basis via an evolution
generated by a time-independent Hermitian operator, iii) define a quantity 
that could probe the spreading while iv) the most efficient spreading is 
quantified by the minimization of the afore-mentioned quantity which is 
equivalent to finding a basis for the state (operator) so that the spreading 
becomes minimum. 

In this section we will develop this general framework for state complexity. 
In the section to follow, we will extend this algorithm to study operator 
complexity.

\paragraph{Complexity of a generic given state} Let us consider a quantum 
system described by a time independent Hamiltonian
whose eigenstates and eigenvalues are denoted by $|E_a\rangle$ and $E_a$, 
respectively. Here $a=1,2,\cdots {\cal D}$ with ${\cal D}$ being the 
dimension of the associated Hilbert space ${\cal H}$. 

For a given Hermitian operator, ${\cal A}:{\cal H}\rightarrow {\cal H}$, one 
can construct an orthonormal and 
ordered basis associated with any state of this Hilbert space. Denoting the 
corresponding 
state by $|\psi\rangle$, the ordered orthonormal basis $\{ |n\rangle, 
n=0,1,2,\cdots, {\cal D}_\psi-1\}$
can be constructed using the Gram-Schmidt process. The first element
of the basis is the given state of the Hilbert state
$|0\rangle=|\psi\rangle$ which we assume to be normalized. Then the other 
elements  are constructed recursively as follows
\be\label{GS}
|\widehat{ n+1}\rangle=({\cal A}-a_n)|n\rangle -b_n|n-1\rangle 
\ee
where $|n\rangle =b_n^{-1}|{\hat n}\rangle$ and
\be
a_n=\langle n|{\cal A}|n\rangle,\;\;\;\;\;\;\;\;\;b_n=
\sqrt{\langle{\hat n}|{\hat n}\rangle}
\ee
This recursive procedure stops whenever $b_n$ vanishes which occurs for 
$n={\cal D}_\psi$ 
defined as the dimension of subspace ${\cal H}_\psi$ expanded by the basis 
$\{ |n\rangle\}$. The dimension of ${\cal H}_\psi$ is in  general smaller 
than the dimension of the full Hilbert space: ${\cal D}_\psi< {\cal D}$.
Note that this procedure produces an orthogonal basis together with 
coefficients $a_n$ and $b_n$ known as the Lanczos coefficients 
\cite{Lanczos:1950zz}.

Since the basis of the subspace ${\cal H}_\psi$ is an ordered basis by 
construction, one can label
any element of the subspace by a number which amounts to defining 
a label operator as 
\be
\ell=\sum_{n=0}^{{\cal D}_\psi-1}\, c_n |n\rangle\langle n|\,,
\ee
 for arbitrary functions $c_n$ which is the ``label'' associated with the 
 state 
 $|n\rangle$. Note that for $n>n'$ one assumes $c_n>c_{n'}$. Since the basis 
 $\{|n\rangle\}$ is already ordered, a natural 
 choice for the coefficient  $c_n$ is $c_n=n$.
 
By construction, the basis $\{|n\rangle\}$ defines a complete basis for the 
subspace 
 ${\cal H}_\psi$. Therefore, any state $|\phi\rangle\in {\cal H}_\psi$ can 
 be expanded as
\be\label{RS}
|\phi\rangle =\sum_{n=0}^{{\cal D}_\psi-1}\,\phi_n|n\rangle,\;\;\;\;{\rm with
}\,\;\;\sum_{n=0}^{{\cal D}_\psi-1}\,|\phi_n|^2=1\,.
\ee 
The expectation value of the label operator in this state $|\phi\rangle$ is 
given by
  \bea
 \langle \phi|\ell|\phi\rangle
={\rm Tr}(\ell\rho_\phi)=\sum_{n=0}^{{\cal D}_\psi-1}\, n|\phi_n|^2 \,,
 \eea
 where $\rho_\phi=|\phi\rangle \langle \phi|$ is the density matrix 
 associated with the state $|\phi\rangle$.
 
Using this expression, one can assign a ``spreading number'' to a  given 
state 
$|\phi\rangle$  in terms of its density matrix as
\be\label{SpN}
{\cal C}_\phi={\rm Tr}(\ell \rho_\phi)\,.
\ee
As its name suggests, \eqref{SpN} can be thought of as a quantity that 
measures the spreading of state $|\phi\rangle$ in the orthogonal basis 
$\{|n\rangle\}$.
Note that, in this notation, the spreading of the state $|m\rangle\in\{|n
\rangle\}$ is $m$. Since the above definition of spreading number is given 
in terms of the density matrix, it can be naturally extended for mixed 
states as well. We will come back to this point later in this paper.

For large ${\cal D}_\psi$ one would expect that for a typical state given by 
\eqref{RS}  with maximum spreading, the spreading is distributed 
statistically over all elements of the basis with equal probability:
$|\phi_n|^2\sim\frac{1}{{\cal D}_\psi}$\cite{Rabinovici:2020ryf}.
From this argument, the maximum value 
for spreading of a state can be estimated as
\be
{\rm Tr}(\ell \rho_\phi)\sim \frac{{\cal D}_\psi}{2}\,.
\ee

Although already evident from our construction, it is worth  emphasizing
that the spreading  number we have associated with the state $|\phi\rangle 
\in {\cal H}_\psi \subset {\cal H}$ depends on two ingredients: the original 
state $|\psi\rangle$ and 
the operator ${\cal A}$ by which the ordered basis is constructed. This is 
very 
reminiscent of the computational complexity in quantum information theory 
\cite{Nielsen} in the sense that  the 
original state $|\psi\rangle$ plays the role of the reference state while 
operator ${\cal A}$, or equivalently the ordered basis $\{ |n\rangle \}$, 
can be thought of as quantum gates. 

Endowed with this interesting identification, we can go ahead to define 
complexity using \eqref{SpN} as follows. After fixing the reference state, 
$|\psi\rangle$, one can construct the 
ordered basis using different Hermitian operators which can be thought of as 
considering 
different gates. If one can find a Hermitian operator  among all possible 
operators, the basis constructed from which minimizes the spreading number
\eqref{SpN}, then the corresponding spreading number we will define as the 
complexity of the state. Therefore, in this context, finding complexity of a 
state boils down to the problem of finding the optimal operator, $\cal 
A_{\rm opt}$.

We note, however, that  in general, for given reference and 
target states, this minimization procedure to find $\cal A_{\rm opt}$ is  a 
complicated program. However, we will show now that for particular cases in 
which the state is obtained by a unitary transformation from the reference 
state, this can be successfully achieved.

\paragraph{Complexity following a unitary evolution} 
So far we considered a typical autonomous state $|\phi\rangle$ without 
knowing apriori whether this state was obtained through any dynamical
process from an initial reference state. We will now focus on the case where 
the desired (target) state is 
obtained from a reference state via a unitary transformation.
In this case it is straightforward to generalize the notion of spreading 
number as follows. 

Let us consider the following state 
\be\label{UA}
|\phi\rangle\equiv |\psi(s)\rangle=U({\tilde{\cal A}},s)|\psi\rangle\,,
\ee
whose evolution in the parameter space $s$ is governed by the Schr\"odinger-
like equation 
\be
\label{eq:Sch}
i\frac{d}{ds}|\psi(s)\rangle={\cal U}({\tilde{\cal A}},s)|\psi(s)\rangle
\ee
with ${\cal U}({\tilde{\cal A}},s)=i \frac{d}{ds}
U({{\tilde{\cal A}}},s)\,U^{-1}({\tilde{\cal A}},s)$.
Here the unitary operator $U$ which evolves the state from an initial 
reference state $|\psi\rangle \equiv |\psi(0)\rangle$, is a function of the 
Hermitian operator ${\tilde{\cal A}}$ and the parameter $s$ which defines 
the flow through \eqref{eq:Sch}. The latter can be chosen such that at $s=0$ 
one has $U=1$. Following its definition given in \eqref{SpN}, the spreading 
number at any arbitrary point  $s$ on the flow is then given by
\be\label{complexity}
{\cal C}(s)={\rm Tr}(\ell \rho(s))=\sum_{n=0}^{{\cal D}_\psi-1}\,n\,
\langle n|\psi(s)\rangle\langle\psi(s)|n\rangle\,,
\ee
where $\rho(s)=|\psi(s)\rangle\langle \psi(s)|$ is the density of state at 
$s$ and the label operator $\ell$ is constructed using a set of orthonormal 
basis $\{|n\rangle\}$ which is complete in ${\cal H}_\psi$.  
This basis can in principle be constructed from any Hermitian operator 
${\cal A}$ acting on the Hilbert space ${\cal H}_\psi$ using the Gram-
Schmidt procedure as in \eqref{GS}. One can expand the target state in the 
same basis, the coefficients of this expansion
being functions of $s$ as
\be
|\psi(s)\rangle=\sum_{n=0}^{{\cal D}_\psi-1}\,\psi_n(s)\,|n\rangle\,,
\ee
which yields, from \eqref{complexity},
\be
{\cal C}(s)=\sum_{n=0}^{{\cal D}_\psi-1}\, n|\psi_n(s)|^2 \,.
\ee
In principle, the coefficients $\psi_n(s)$ can be recursively read off from 
the equation \eqref{UA} if one is equipped the full knowledge of Lanczos 
coefficients $a_n$ and $b_n$.


Let us now denote the eigentstates and eigenvalues of the operator 
${\tilde{\cal A}}$ by
$\{|\alpha_i\rangle\}$ \footnote{$\{|\alpha_i\rangle\}$ do not in general 
form an ordered basis.} and $\alpha_i$, respectively. Since ${\tilde{\cal 
A}}$ is Hermitian, these eigenstates form a complete set of states in ${\cal 
H}_\psi$. Using this fact, the spreading number \eqref{complexity} can be 
recast into the following form
\be
{\cal C}(s)=\sum_{\alpha_1,\alpha_2}\,U(\alpha_1,s)U^*(\alpha_2,s)
\rho_0(\alpha_1,\alpha_2)  \langle \alpha_1|\ell|\alpha_2\rangle\,,
\ee
where $\rho_0(\alpha_1,\alpha_2)=\langle\alpha_1|\psi(0)\rangle\langle 
\psi(0)|
\alpha_2\rangle$ is the density matrix in the $\alpha$-basis. $\langle
\alpha_1|\ell|\alpha_2\rangle$ are the matrix elements of the label operator 
in  eigenvectors of the Hermitian operator
$\tilde{\cal A}$ given by
 \be
 \label{eq:l-exp1}
 \langle \alpha_1|\ell|\alpha_2\rangle=\sum_{n=0}^{{\cal D}_\psi}\,n
 \langle\alpha_1|n\rangle\langle n|\alpha_2\rangle  \,.
 \ee

It is worth noting here that these matrix elements can be computed directly 
in the continuum limit using the recursion relation given in \eqref{GS}. 
Expanding $|n\rangle$ in the basis of eigenvectors of the operator 
${\tilde{\cal A}}$
 \be
 |n\rangle =\sum_i c_n(\alpha_i)|\alpha_i\rangle\,,
 \ee
the equation \eqref{GS} reads
 \be
 \label{eq:GS-alpha}
 \alpha c_n(\alpha)= a_nc_n(\alpha)+b_nc_{n-1}(\alpha)+b_{n+1}c_{n+1}
 (\alpha),
 \ee
 where $c_n(\alpha)=\langle \alpha|n\rangle$, $|\alpha\rangle$ being a
 particular eigenstate of $\tilde{\cal A}$ with non-degenerate eigenvalue 
 $\alpha$. This equation can be thought of 
 as a time independent Schr\"odinger equation for which one can find the 
 wave functions $c_n(\alpha)$ recursively. Using  these wave functions, one 
 can further compute the matrix elements  \eqref{eq:l-exp1}.
 
In order to obtain an expression for the matrix elements in the continuum 
limit, one can first rescale   $c_n(\alpha)\rightarrow
 (i)^nc_n(\alpha)$ and then set $x=n\epsilon,  \,b(x)=2\epsilon
 b_n, a(x)=a_n$ and $c_n(\alpha)=c(x,\alpha)$. Thereafter, expanding 
 \eqref{eq:GS-alpha} upto the leading order in $\epsilon$ yields a much 
 simpler equation
\be
\label{eq:GS-cont}
-i(\alpha- a(y))f(y,\alpha)=\partial_yf(y,\alpha),
\ee
where $\partial_y=b(x)\partial_x$ and $f(y,\alpha)=\sqrt{b(x)}c(x,\alpha)$. 
The equation \eqref{eq:GS-cont} can be readily solved to obtain
\be
f(y,\alpha)=f(0,\alpha)\,e^{-i\alpha y+ i\int_0^y a(y')dy'}\,.
\ee

Using this solution, finally one arrives at an expression for the matrix 
element  \eqref{eq:l-exp1} as
\be\label{MEL}
\langle \alpha_1|\ell|\alpha_2\rangle=\frac{1}{\epsilon^2}\int
dy\, x(y)\,e^{-i\alpha_{12}y}
\ee
where $\alpha_{12}=\alpha_1-\alpha_2$.
As its form suggests, to evaluate the matrix elements explicitly, one needs 
to know $x$ as a function of $y$ which 
can be obtained from the relation $\frac{dx}{b(x)}=dy$. Given $b(x)$, the 
integral can 
be performed to find $x(y)$ and subsequently, the explicit expression for 
the desired matrix elements which determine the spread in the continuum 
limit.

\paragraph{Minimising the spreading and complexity:} Since the flow itself 
is generated in terms of a particular Hermitian operator $\tilde{\cal A}$, 
it is natural to consider a special case when ${\tilde{\cal A}} = {\cal A}$ 
that is when ordered basis is constructed using the same Hermitian operator 
that generates
the unitary flow \eqref{UA}.
One can show that in this case the spreading number is minimum compared to 
any other choice of Hermitian operator or equivalently, for any other choice 
of basis. This follows directly from theorem 1 of 
\cite{Balasubramanian:2022tpr}, by setting $n = 1$.

Therefore the resultant spreading number obtained with the choice of 
${\tilde{\cal A}} = {\cal A}$ can be
considered as the complexity of the s-evolved target state. Indeed, looking 
at equation \eqref{complexity}
one finds that the minimum of this with $\{|n\rangle\}$ being the 
orthonormal basis generated by the Hermitian operator ${\cal A}$ is of the 
same form as that of the Krylov complexity \cite{{Parker:2018yvk},
{Barbon:2019wsy},{Jian:2020qpp},{Rabinovici:2020ryf}} (see also 
\cite{Fan:2022xaa,Muck:2022xfc,Fan:2022mdw}).

It is worth mentioning that  due to 
the property of the trace, the complexity defined above at given $s$ may 
also be  given by
\be
{\cal C}(s)={\rm Tr}(\ell(s) \rho),
\ee
that means, we keep the initial reference state fixed and  construct the 
ordered basis
using the Hermitian operator ${\cal A}$ appearing in the evolution operator 
${\cal U}({\cal A},s)$.

\paragraph{The time-evolution as an example:}
As a particularly interesting  and rather physical example of the 
generalized framework developed above, let us  consider the case where 
the Hermitian operator ${\cal A}$ is the Hamiltonian of the quantum system. 
This is the example mainly studied in the literature to compute 
Krylov complexity. In this case, starting with an initial state $|\psi(0)
\rangle$, 
the dynamics is given by the Schr\"odinger equation
\be\label{St1}
|\psi(t)\rangle =e^{iHt} |\psi(0)\rangle\,.
\ee
Then the  density matrix associated with this state at any time $t$ is given 
by
\be
\rho(t)=|\psi(t)\rangle\langle\psi(t)|
=e^{iHt}\,\rho(0)\, e^{-iHt}
\ee
where $\rho(0)=|\psi(0)\rangle\langle\psi(0)|$, by which the complexity is
\be\label{KC}
{\cal C}(t)={\rm Tr}(\ell\rho(t))=\sum_{n=0}^{{\cal D}_\psi-1}\,n\,|
\psi_n(t)|^2\,,
\ee
where $\psi_n(t)$ is given by the following expression where the state is  
expanded in terms of the ordered basis $\{|n\rangle\}$
(known as Krylov basis in this case)
\be
|\psi(t)\rangle=\sum_{n=0}^{{\cal D}_\psi-1}\,\psi_n(t)|n\rangle, 
\ee
Using the Schr\"odinger equation, one can deduce an equation for $\psi_n(t)$ 
that can be solved recursively using the knowledge of Lanczos coefficients 
$a_n$ and $b_n$.
Actually for a chaotic system these coefficients, as functions of $n$, 
follow a universal behavior that includes a linear growth 
\cite{Parker:2018yvk}  followed by a saturation \cite{{Barbon:2019wsy},
{Jian:2020qpp},{Rabinovici:2020ryf}}.
 These behaviors give an early-time exponential and the late times linear 
 growths for complexity respectively 
 \cite{{Barbon:2019wsy},{Jian:2020qpp},{Rabinovici:2020ryf}}.

In order to proceed further to explore the complexity associated with the 
Hamiltonian evolution, let us assume that the Hamiltonian
of the system has a continuous spectrum. Therefore, using the energy 
eigenstates, 
the Krylov complexity \eqref{KC} may be recast into the following form  
\be\label{CInt}
{\cal C}(t)=\int dE_a\,dE_b\,e^{i(E_a-E_b)t}\rho_{0}(E_a,E_b)\,A(E_a,
E_b)\,,
\ee
where 
\bea
&&\rho_{0}(E_a, E_b)=\langle E_a|\rho(0)|E_b\rangle,\;\;\;\;A(E_a,E_b)=
\langle E_a|
\ell|E_b\rangle\,, \nonumber\\
\eea
which is the same expression proposed for complexity in 
\cite{Alishahiha:2022nhe}. In \cite{Alishahiha:2022nhe}, the left hand side 
of \eqref{CInt} was identified as complexity provided the function 
$A(E_a,E_b)$ had a double pole structure at late-time, i.e. in the 
coincident limit $E_a \rightarrow E_b$, thereby restricting the function 
$A(E_a,E_b)$ to a class of quantum expectation values which violates the 
ETH. We shall argue below that this is naturally the case when we consider 
the expectation value of the label operator which turns out to be the 
atypical non-local operator existence whereof was postulated in 
\cite{Alishahiha:2022nhe}. 

As we have already mentioned the saturation of Lanczos results in a linear 
growth at late times. In the notation of \eqref{CInt} this behavior at  late 
times imposes
a condition on the function $A(E_a,E_b)$ to have a  double pole structure 
\be
A(E_a,E_b)=-\frac{a(E)}{\omega^2}+{\rm local \,
terms},\;\;\;\;\;\;\;\;{\rm for}\,\,\omega\rightarrow 0\,,
\ee
 where $\omega=E_a-E_b$ and $2E=E_a+E_b$, in agreement with 
 the proposal of \cite{Alishahiha:2022nhe}. We can derive the surprising 
 connection between 
 the double pole structure of the function $A(E_a,E_b)$ and the saturation 
 of Lanczos coefficients from the expectation value of the label operator in 
 the continuous limit given in 
 \eqref{MEL}. When the Lanczos coefficient $b(x)$ saturates to a constant, 
 one gets $x=y$. This yields, from the expression  \eqref{MEL},
\be
\langle E_a|\ell|E_b\rangle=\frac{1}{\epsilon^2}\int_0^\Lambda
dy\, x(y)\,e^{-i\omega y}=-\frac{1}{\omega^2}\,(1-e^{-i\omega\Lambda})
\ee
where $\Lambda$ is a cutoff. This is the double pole structure we expect to 
get 
from $A$-function that generates linear growth for complexity at late time.
For $\Lambda\rightarrow \infty$ the above integral may be recast into the 
following form 
\be
\langle E_a|\ell|E_b\rangle=\frac{1}{\epsilon^2}\int_0^\infty
dy\, x(y)\,e^{-i\omega y}=\frac{1}{\epsilon^2}\,i\frac{d}{d\omega}
\delta(\omega)\,.
\ee
On the other hand, in the regime when the Lanczos coefficient $b(x)$ 
exhibits a linear growth, say, $b(x)=2\lambda x$, one obtains the relation 
$x=e^{2\lambda y}$, which upon inserting in equation \eqref{MEL} yields
\be\label{EE}
\langle E_a|\ell|E_b\rangle=\frac{1}{\epsilon^2}\delta(\omega+2i\lambda)\,.
\ee
Plugging this 
result in \eqref{CInt} one finds early 
time exponential growth ${\cal C}(t)\sim e^{2\lambda t}$ as the case for 
chaotic system.

For a more general case of $x=y^m$ for $m>1$, one gets
\be
\langle E_a|\ell|E_b\rangle=\frac{1}{\epsilon^2}\,i^m\frac{d^m}{d\omega^m}
\delta(\omega)\, ,
\ee
which can happen at early times in a non-chaotic system. Therefore, from 
\eqref{CInt} one gets ${\cal C}(t)\sim t^m$ which can be interpreted as the 
early time power-law growth for complexity.

An advantage to study the matrix elements of the label operator is that the 
universal 
behavior of Lanczos coefficients associated with a Hermitian operator may be 
studied without refereeing to a particular dynamics by which a state 
evolves.

To explore the significance of the label operator better,
 we note 
 that for an arbitrary operator $\Lambda=\sum_a \lambda_a|\lambda_a\rangle
 \langle \lambda_a|$
 where  $\lambda_a$ and $|\lambda_a\rangle$ are its eigenvalues and 
 eigenstates,
 respectively, one can compute ${\rm Tr}(\Lambda \rho(t))$ which has the 
 same form as that of
 \eqref{CInt}, though in this case it is not guaranteed that the functions 
 $A(E_a,E_b)=\langle E_a|\Lambda |E_b\rangle$
  exhibit double pole structures at late times. Actually we would expect 
  that for a typical 
  operator $\Lambda$ this satisfies the ETH ansatz and thus 
  ${\rm Tr}(\Lambda \rho(t))$ gives a time independent quantity interpreted 
  as the average 
  of the operator $\Lambda$. Therefore in order to have the notion of 
  complexity 
  it is important to compute \eqref{CInt} specifically for the label 
  operator of an ordered basis. { This justifies further the uniqueness 
  of our definition of complexity in terms of the label operator.}
     
 The other important element of the formula given in \eqref{CInt} is the 
 density matrix $\rho_{0}(E_a,E_b)$.
 At the leading order in the dimension of the 
 Hilbert space, the density matrix $\rho_{0}(E_a,E_b)$ is factorized,
 though in general it has a form
\be
\label{eq:rhoexp}
\rho_{0}(E_a,E_b)= \rho(E_a)\rho(E_b)+\rho_{c}(E_a,E_b)\,.
\ee
Here $\rho_{\rm c}$ represents the connected term meaning that it cannot be 
written in a factorized form of $g_1(E_1)g_2(E_2)$ with $g_{1,2}$ being 
arbitrary functions of energy. While this object remains a silent spectator 
in the discussion of the late time linear growth of complexity above, we 
will now argue that it plays a pivotal role in understanding saturation 
phase of complexity at later times.

\paragraph{On the saturation of complexity at late time:} In what follows we 
would like to present a general form of complexity in 
$\tau$ scaling limit in which we take $\{t,{\cal D}_\psi\} \rightarrow  
\infty$
 while keeping $\tau=t \, {\cal D}_\psi^{-1}$ fixed.
 
We proceed by rescaling $\rho_0(E_a,E_b)={\cal D}_\psi^2
\tilde{\rho}_0(E_a,E_b)$ and  switching to the $(E,\omega)$ coordinates. The 
latter is convenient for studying the coincident limit $E_a \rightarrow E_b$ 
relevant for the late time limit $t \rightarrow \infty$. With this, the 
complexity \eqref{CInt}, up to an appropriate normalization, reads
 \be
{\cal C}(t)=
\int_0^\infty dE\int_{-\infty}^\infty d\omega\,
e^{i\omega t}\tilde{\rho}_{0}
(E,\omega) A(E,\omega)\,.
\ee
 As we have already demonstrated above, the saturation phase of the Lanczos 
 coefficients, 
 which corresponds to the $\tau$ scaling limit,
 results in a double pole structure for $A(E,\omega)$. Assuming to have an 
 expression
 for complexity growth at leading order consistent with Lloyld's bound one 
 arrives at \cite{Alishahiha:2022nhe}
 \be
  A(E,\omega)=-\frac{\sqrt{E}}{\tilde{\rho}(E)\,\omega^2}+{\rm Local\, 
  terms}\,.
  \ee
On the other hand, at the $\tau$ scaling limit, the physics is dominated by 
correlations
between nearby energy levels. In this limit, the connected part of the 
matrix elements of 
the density  matrix denoted by $\rho_c(E_a, E_b)$ in \eqref{eq:rhoexp}, are 
described by the universal sine-kernel
formula \cite{Mehta}. More precisely, one has
\be
\tilde{\rho}_0(E,\omega)=\tilde{\rho}(E)^2+\frac{{\tilde \rho}(E)}
{{\cal D}_\psi}
\delta(\omega)-\frac{\sin^2({\cal D}_\psi {\tilde \rho}(E)\omega)}{({\cal
D}_\psi\omega)^2}\,.
\ee 
Putting everything together, one arrives at
\bea
&&{\cal C}(t)=C_0-
{\cal D}^{-1}_\psi\int_0^\infty dE \sqrt{E}\int_{-\infty}^\infty d\omega\,
\frac{e^{i\omega t}}{\omega^2}\,
\delta(\omega) \cr &&\cr 
&&\;\;\;\;\;\;\;\;\;\;\;\;-
\int_0^\infty dE \sqrt{E}\tilde{\rho}(E)\int_{-\infty}^\infty d\omega\,
\frac{e^{i\omega t}}{\omega^2}\cr &&\cr&&\;\;\;\;\;\;\;\;\;\;\;\;\;\;
\;\;\;\;\;\;\;\;\;\;\;\;\;
\times\left(1-\frac{\sin^2({\cal D}_\psi {\tilde \rho}(E)\omega)}
{({\cal D}_\psi\tilde{\rho}(E)\omega)^2}\right),
\eea
where $C_0$ is a constant. It is exactly the same expression which was 
obtained 
for (super)JT gravity in which the density $\tilde \rho$ is known explicitly
\cite{{Iliesiu:2021ari},{Alishahiha:2022kzc},{Alishahiha:2022exn}}.
In this equation, the first line which is divergent in general and needs to be 
regularized by a cutoff, is nevertheless time independent and does not contribute at late times. Therefore, in what 
follows we only need to consider the last term. Setting $\omega t=\xi$, this term yields
\bea
&&{\cal C}(t)=-{\cal D}_\psi\,\tau\int_0^\infty dE \sqrt{E}\tilde{\rho}(E)\int_{-\infty}^\infty d\xi\,
\frac{e^{i\xi}}{\xi^2}\cr &&\cr &&\;\;\;\;\;\;\;\;\;\;\;\;\;\;
\;\;\;\;\;\;\;\;\;\;\;\;\;\;\;\;\;\;\;
\times\left(1-\frac{\sin^2(\frac{{\tilde \rho}(E)}{\tau}\,\xi)}
{(\frac{{\tilde \rho}(E)}{\tau}\,\xi)^2}\right),
\eea
which vanishes for $\tau>\tilde \rho(E)$ for any density function $\tilde \rho$
\cite{Iliesiu:2021ari}. This behaviour is quite universal in the sense that the saturation phase  does not depend on the details of the model.
On the other hand, for $\tau<\tilde \rho(E)$  one can expand the sine 
function in terms of exponential functions and performing the complex integral over a contour excluding the poles in the lower half plane one obtains \cite{{Iliesiu:2021ari},{Alishahiha:2022kzc}}
\bea
{\cal C}(t)\approx-\frac{2\pi{\cal D}_\psi}{3}\,\int_{E_\tau}^\infty dE 
\sqrt{E}\,\tilde\rho^2(E)
\left(1-\frac{\tau}{2\tilde\rho(E)}\right)^3 \;\;\;
\eea
where $E_\tau$ can be read off from the equation $\tau=\tilde \rho(E_\tau)$. 
Although 
to perform the integration over $E$ one needs to know the explicit form of 
the 
density $\tilde\rho$, one can still extract certain universal behavior for 
the complexity.
In particular for $\tau\ll 1$ one gets ${\cal C}(t)\sim -\alpha_0 
{\cal D}_\psi+\alpha_1 t$ with $\alpha_{0,1}$ being order one constants. 
Although the saturation 
phase occurs at $\tau\sim 1$ limit is also universal, the actual way the 
complexity approaches the saturation phase is model dependent and 
is fixed as soon as the density is fixed. 

To conclude, we note that for a chaotic system, the complexity has a
universal behavior starting with early times exponential growth and linear 
growth
at  late times followed by a saturation. It is worth noting that while 
the
early exponential and the late times linear growths are described by the 
behavior of the 
$A(E,\omega)$ function, already at leading order, the saturation phase is 
the consequence  of the contribution of the connected part of the 
density-density short range correlation known as sine-kernel.

This has to be compared with the numerical computations done {\it e.g.} in \cite{Rabinovici:2022beu},
where the saturation of complexity was due to the descent phase of the Lanczos coefficients. 
It would be interesting to understand and compare these two different approaches by which the complexity reaches the saturation phase. An early saturation of complexity due to the breaking of ETH might actually signal the chaotic nature of the system. Work in this direction is in progress \cite{Inprogress} and we hope to report on this soon.

 \section{Complexity: The Operator-State correspondence}
 \label{sec3}

 In this section we will extend the algorithm elaborated in the previous 
 section, to the study of complexity corresponding to growth of an operator. 
 To achieve this, generally, one starts with a reference 
 operator and constructs an ordered basis of operators using a Hermitian 
 operator. Then 
 one looks for the spreading of the desired operator over this ordered  
 basis.
 
Typically, in order to go through this procedure, one needs to define
a proper inner product in the space of operators. However, in what follows, 
we will take a different approach so that the inner product will arise 
naturally. 
 
\paragraph{The doubled Hilbert space:} To compute complexity associated with 
the growth of an operator we will essentially use the same procedure as that 
of the state complexity simply by making use of  the channel-state map 
\cite{Nielsen} which maps an operator to a state. 

Let us consider a quantum system described by the Hamiltonian $H$ with 
eigenstates and
 eigenvalues $|E_a\rangle$ and $E_a$ respectively. The corresponding Hilbert 
 space is also denoted by ${\cal H}$ with dimension ${\cal D}$. 
 In this set up, we will consider a generic operator ${\cal O}:
 {\cal H}\rightarrow 
 {\cal H}$ and a complete basis spanning the Hilbert space, denoted by 
 $\{|i\rangle,
 i=1,2,\cdots {\cal D}\}$. The matrix elements of the operator in this basis 
 is
 ${\cal O}_{ij}=\langle i|{\cal O}|j\rangle$. Following the channel-state 
 duality, one can then define an associated
 state, through a linear bijection \footnote{This is known in literature as 
 the Choi-Jamiolkowski isomorphism \cite{PhysRevA.87.022310}.}, in an 
 auxiliary doubled Hilbert space ${\cal H}_d={\cal H}\otimes {\cal H}$ as 
 follows
\be\label{Eq1}
|\psi_{\cal O}\rangle=\sum_{i,j=1}^N \varrho_{ij}\,|i\rangle\otimes | 
j\rangle
\equiv \sum_{i,j=1}^N \varrho_{ij}\,|i,j\rangle\,.
\ee
Here the density of state $\varrho_{ij}$ is proportional to the matrix 
elements ${\cal
O}_{ij}$ such that $\sum_{i,j} |\varrho_{ij}|^2=1$. More precisely, one has
\be
\varrho_{ij}=\frac{{\cal O}_{ij}}{\sqrt{\sum_{i',j'}{\cal O}_{i'j'}{\cal 
O}_{i'j'}}}\;.
\ee
This map would naturally define an inner product between two operators
\be
{\cal O}_1\cdot {\cal O}_2\equiv \langle \psi_{{\cal O}_1} |\psi_{{\cal 
O}_2}\rangle.
\ee
Once we have the state in the doubled Hilbert space, we can simply adopt our 
previously developed formalism for state complexity, now for a state in the 
doubled Hilbert space. For a given Hermitian operator 
${\cal A}:{\cal H}\rightarrow {\cal H}$ one can then readily construct an 
orthonormal and
ordered basis in the doubled Hilbert space with the identification of the 
first state 
as $|0,0\rangle=|\psi_{\cal O}\rangle$. All other elements are obtained 
using the algorithm given in \eqref{GS}, though, now in the doubled Hilbert 
space
\be
\label{eq:double-ordered}
|\widehat{ n+1},\widehat{ n+1}\rangle=({\cal A}_d-a_n)|n,n\rangle -b_n|n-
1,n-1\rangle 
\ee
where $|n,n\rangle =b_n^{-1}|{\hat n,\hat n}\rangle$ and
\be
a_n=\langle n,n|{\cal A}_d|n,n\rangle,\;\;\;\;\;\;\;\;\;b_n=
\sqrt{\langle{\hat n},\hat n
|{\hat n},\hat n\rangle}\,.
\ee
where ${\cal A}_d={\cal A}\otimes {\cal A}$ is an operator acting on the 
doubled Hilbert space ${\cal A}_d:{\cal H}_d\rightarrow {\cal H}_d$.
 The resulting ordered
basis defines a subspace of the doubled Hilbert space denoted by 
${\cal H}_{d,\psi_{\cal O}}$, ${\cal D}_{\psi_{\cal O}}$ being the dimension 
of this sub Hilbert space. 

For a given operator ${\cal O}$, the spreading is now defined as
\be
\label{eq:spreading-doubled}
{\cal C}_{{\cal O}}={\rm Tr}(\ell \rho_{{\cal O}})
\ee
where $\ell=\sum_{n=0}^{{\cal D}_{\psi_{\cal O}}}n|n,n\rangle\langle n,n|$ 
and $\rho_{\cal O}=|\psi_{\cal O}\rangle\langle \psi_{\cal O}|$ is the 
density matrix of the state $ |\psi_{\cal O}\rangle$ associated with the 
operator ${\cal O}$.

Following this definition, it is then a straightforward  task to compute 
complexity of an operator
that is obtained from a reference operator ${\cal O}$ via a unitary 
evolution. Let us consider the following unitary transformation for an 
operator
\be
{\cal O}(s)={\cal U}({\cal A},s)\; {\cal O}\;{\cal U}^{-1}({\cal A},s)
\ee
so that
\be
\rho_{\cal O}(s)=\sum_{i,j=1}^{{\cal D}}\sum_{i'j'=1}^{\cal D}\;
\varrho_{ij}(s)\varrho^*_{i'j'}(s)\;
|i,j\rangle\langle j',i'|
\ee
where $\varrho_{ij}(s)=\langle i|{\cal O}(s)|j\rangle/\sqrt{\sum |
{\cal O}_{ij}|^2}$.

From this, using the definition of spreading in 
\eqref{eq:spreading-doubled}, one gets
\be
\label{complexity-doubled}
{\cal C}(s)={\rm Tr}(\ell \rho_{{\cal O}}(s))=
\sum_{n=0}^{{\cal D}_{\psi_{\cal O}}-1}n |{\cal O}_n(s)|^2
\ee 
where 
\be
{\cal O}_n(s)=\sum_{i,j}^{{\cal D}}\varrho_{ij}(s)\langle n,n|i,j\rangle\,
\ee
which implies to have the following expansion for the state in terms of 
ordered basis in the doubled Hilbert space
\be
|\psi_{\cal O}(s)\rangle=\sum_{n=0}^{{\cal D}_{\psi_{\cal O}}-1}
{\cal O}_n(s)\,|n,n\rangle\,.
\ee
One can also take the basis $\{|i\rangle\}$ to be the eigenstates of the 
Hermitian operator ${\cal A}$. Note that, here we considered the evolution 
of the operator by the same operator $\cal A$ used to construct the ordered 
basis in the doubled Hilbert space. Following our argument in the case of 
state complexity, this ensures that \eqref{complexity-doubled} represents 
minimum spreading and therefore can be interpreted as the complexity of the 
target operator ${\cal O}_s$.

\paragraph{The time evolution revisited:} As before, our main interest lies 
in the study of growth of an operator following  the evolution with the 
Hamiltonian of 
the system
\be\label{OG}
 {\cal O}(t)=e^{iHt}{\cal O}e^{-iHt}\,.
 \ee
In this case, it is natural to use the energy eigenstates for the channel-
state map
by which the corresponding reference space is given by \eqref{Eq1} with the 
replacement $|i,j\rangle\rightarrow |E_a,E_b\rangle$ 
\be\label{OSt}
|\psi_{\cal O}\rangle=\sum_{a,b}^{\cal D} \varrho_{ab}\,|E_a,E_b\rangle\,,
\ee
where $\varrho_{ab}$ is the normalized matrix elements of the operator
in the energy basis. At a given time $t$, the corresponding matrix elements 
are
\be
\varrho_{ab}(t)=
e^{i(E_a-E_b)t}\varrho_{ab},
\ee
so that the associated state in doubled Hilbert space is transformed as 
follows
\be\label{StEv}
|\psi_{\cal O}(t)\rangle=e^{iH_-t}\,|\psi_{\cal O}\rangle
\ee
where $H_-=H\otimes 1-1\otimes H$. In order to compute the complexity 
through minimizing the spreading, following our earlier discussion, one 
needs to construct the ordered basis, specifically using 
the Hermitian operator ${\cal A}=H_-$. In this case, the general 
algorithm mentioned above, yields the operator complexity
\be
{\cal C}(t)={\rm Tr}(\ell \rho_{{\cal O}}(t))=
\sum_{n=0}^{{\cal D}_{\psi_{\cal O}}-1}n |{\cal O}_n(t)|^2\,,
\ee 
where ${\cal O}_n(t)=
\sum_{a,b}\varrho_{ab}(t) \langle n,n|E_a,E_b\rangle$.

In the case when the Hamiltonian possesses a continuum spectrum, one can 
write down an expression for complexity similar to \eqref{CInt} in the 
doubled Hilbert space. This will lead to a late time saturation of 
complexity followed by a linear growth. The saturation will be guaranteed 
through the perturbative contact terms and non-perturbative contributions 
appearing in the density correlation $\langle E_a, E_{a'}| \rho_{{\cal 
O}}(t))| E_b, E_{b'}\rangle$ in the coincident limits $E_a - E_{b} 
\rightarrow 0$ and $E_{a'} - E_{b'} \rightarrow 0$.

To summarize, we note here that in order to study operator complexity for an 
operator 
evolving with the Heisenberg equation, one can equivalently run the 
algorithm developed to compute the state complexity, but in the doubled 
Hilbert space with the Hamiltonian $H_-$.

It is worth noting that, in this case, the ordered basis, generated using 
the particular effective Hamiltonian $H_-$ which governs the time evolution 
of the state, renders the Lanczos coefficients $a_n$ to be zero. This is of 
course expected due to the fact that the evolution of an operator here is 
given by the Liouvillian which amounts to have vanishing $a_n$. 

Let us now discuss one interesting subtlety of this construction. For 
quantum system defined 
on the  doubled Hilbert space ${\cal H}_d$ whose dynamics is given by $H_-$, 
the average energy $H_+= H\otimes 1+1\otimes H$ is a conserved charge.
This follows from the fact that $[H_-,H_+]=0$. In other words, for an 
operator whose associated 
state is defined by \eqref{OSt}, one  should impose the following condition
 \be
 H_+|\psi_{\cal O}(t)\rangle=E|\psi_{\cal O}(t)\rangle,
 \ee
 where $E=E_a+E_b$ is the average energy which is kept fixed.

The average energy should remain
constant during the state evolution \eqref{StEv} and therefore it does not 
mix states with different energies. In other words the complexity can be 
computed for each individual sector with fixed average energy 
\cite{Kar:2021nbm}.

 Precisely due to this particular dynamics of 
the operator growth, the diagonal part of the matrix elements 
${\cal O}_{ab}$ does
 not contribute to the operator growth. In other words, restricting 
 ourselves to an 
operator whose corresponding state in doubled Hilbert space is given by
\be
|\psi_{\cal O}^{(0)}\rangle=\sum_{a}\varrho_{a}|E_a,E_a\rangle
\ee
one gets trivial dynamics under a unitary time evolution given by $H_-$. 
Here the 
index $(0)$ indicates that this state belongs to a subspace with constant 
$H_-$ (which could be zero).

Actually this defines rather atypical states whose dynamics are, rather 
naturally, given by 
$H_+$,
\be
\label{eq:TFD1}
 |\psi_{\cal O}^{(0)}(t)\rangle =e^{i H_+t}|\psi_{\cal O}^{(0)}\rangle\,.
\ee
Note that for $\varrho_{a}=e^{-\beta E_a/2}/\sqrt{\sum_a e^{-\beta E_a}}$
 this state corresponds to a
thermofield double state with inverse temperature $\beta$. In general 
$\varrho_a$ 
could be a complex function having a phase $\varrho_a=|\varrho_a|
e^{i\alpha_a}$ which could be thought of the generalized thermofield 
double state \cite{Verlinde:2020upt,Papadodimas:2015xma}.

This apparently confusing role reversal of $H_\pm$, mentioned above in the 
context of the atypical state, is actually not that surprising. In fact, in 
the construction of the doubling of the Hilbert space in \eqref{OSt}, it is 
assumed that energy eigenstates of the Hamiltonians participating in the 
doubling are also eigenstates of the time reversal operators, as is the case 
of most quantum mechanical system. This is, however, not generically true. 
Taking this into account, explicitly, \eqref{OSt} can be rewritten as 
\be\label{OSt-general}
|\psi_{\cal O}\rangle=\sum_{a,b}^{\cal D} \varrho_{ab}\,
{\cal T}|E_a,E_b\rangle\,,
\ee
where ${\cal T}$ is the time-reversal operator, which being an anti-linear 
operator, results in the normalized matrix elements 

\be
\varrho_{ab}(t)=
e^{i(E_a + E_b)t}\varrho_{ab}.
\ee 
Then the associated evolution of the state in the doubled Hilbert space is 
given by
\be\label{StEv-general}
|\psi_{\cal O}(t)\rangle=e^{iH_+t}\,|\psi_{\cal O}\rangle\,.
\ee
When $|\psi_{\cal O}\rangle$ is the TFD state, such time evolutions give 
rise to phase-shifted thermofield doubled states mentioned above. 

Therefore, we note that there are two different ways to arrive at the same 
generalized TFD states, either as a simple one-sided time evolution 
\eqref{eq:TFD1}, preserving the time-reversal symmetry of the basis states, 
or in terms of the doubled Hilbert space structure using energy eigenstates 
which are not time-reversal symmetric as we derived in \eqref{StEv-general}. 

These two points of view correspond to the boundary and bulk perspectives, 
respectively, in the context of AdS/CFT.  The former approach remains 
faithful to the time-reversal symmetry of the CFT while creating an enlarged 
phase space of complexified quantum states with asymptotic charges 
\cite{Harlow:2015lma}, and the latter creates a complexified bulk states 
through evolution with $H_+$ identified as the bulk Hamiltonian. 
Accordingly, the additional phase appearing in the generalized TFD state 
assumes interpretations either due to boundary charges or as topological 
quantum phases arising as a consequence of bulk non-locality.

This non-locality is quantified through the symplectic structure in the 
bulk, which although can have local interpretation of time evolution with 
$H_+$, are globally non-exact, thereby giving rise to the additional phases 
appearing in the generalized TFD states \cite{Nogueira:2021ngh, 
Banerjee:2022jnv}. 

\paragraph{A little more on doubling and the holographic interpretation:}
Our construction of doubled basis follows a secret rendition of the 
Reeh-Schlieder-Theorem. 
By construction, the domain ${ \Gamma}_\psi$ of any initial given state 
$|\psi\rangle$ is dense in the full Hilbert space of the theory once we are 
successfully able to construct the ordered basis using an algebra of 
operators $\Omega$ that minimize the spreading. Therefore, this state can be 
dubbed as a cyclic vector. Furthermore, to have a non-vanishing complexity, 
one needs to start with a state which is not an eigenstate of the operator, 
which means there cannot be an annihilation operator in the small algebra of 
operators, ${\Omega}$. This makes the initial state $|\psi\rangle$ a 
separating vector. 

With these two defining conditions of the Reeh-Schlieder Theorem 
\cite{Haag:1992hx}, one can visualize the full algebra of operators acting 
on the Hilbert space $\Gamma_\psi$ as an entangled algebra ${\Omega}_d={ 
\Omega}\otimes {\bar{\Omega}}$ where ${\bar{\Omega}}$ is the commutant 
algebra which can be obtained using modular automorphism via the Tomita-
Takesaki theorem \cite{Haag:1992hx}.  Furthermore, given the emerging 
structure of entanglement, it is natural to invoke a doubled basis as in 
\eqref{Eq1} to expand the state and consequently, a doubled ordered basis as 
in \eqref{eq:double-ordered} to analyse the growth of the state.

This justifies why the doubling, introduced in \eqref{Eq1} using the channel-
state map, is essential to cast the operator complexity as a corresponding 
state complexity\footnote{The realisation we achieve through this 
unification is quite in line with the Gelfand-Naimark-Segal (GNS) 
construction discussed in the context of studying operator growths in large 
$\cal N$ theories \cite{Magan:2020iac}.  }. 

In the context of holography, such constructions were instrumental in 
understanding the state-dependent reconstruction of black hole interior 
\cite{Papadodimas:2013jku}\footnote{However, in this case the small algebras 
were only approximate algebra at large $\cal N$ with edge effects which were 
important to realize the existence of the black hole interior. In the 
language of Lanczos, it would corresponding to only approximate breaking of 
the algorithm when $b_n$ becomes smaller to a given hierarchy scale, as 
$\cal N$ is, in the case of holographic CFT's.}. We note here, following our 
unified interpretation of the state and the operator complexities, that the 
same construction is also pivotal to connect to the the notion of 
holographic complexity which measures the growing volume of the interior of 
a black hole. 

It is worth mentioning that in the holographic context, there can actually 
be an infinitely many quantities which possess the same late time behaviour 
in terms of a linear growth followed by a saturation and in principle, all 
of them can be dubbed as complexities \cite{Alishahiha:2022nhe, 
Belin:2021bga}. Identifying the $A$-functions in terms of the matrix elements 
of the label operator potentially removes this ambiguity.
Therefore, the interpretation of both, in terms of operator growth, should 
follow from the same doubling algorithm.

An alert reader might wonder about the identification of the matrix elements 
of the label operator with the expectation values of a non-local holographic 
position operator. However, we would like to emphasize that the entanglement 
between two spacelike separated regions can have an equivalent description 
in terms of entangling algebras of operators defined on the Hilbert space. 
While the previous description is more geometric and visually pleasing, the 
latter one provides a more general notion of entanglement independent of 
spacetime. Following the interpretation of the doubled Hilbert space in 
terms of the modular automorphism mentioned above, this generalized notion 
of entanglement justifies the matrix elements of the label operator being 
identified with the holographic non-local operator. {More details regarding this identification will be elaborated in \cite{Inprogress}.}

  \section{Subregion Krylov complexity}
  \label{sec4}
  
 Entanglement entropy or other measures of entanglements are given in terms 
 of the reduced density matrix. This is unlike the complexity the usually is 
 defined for a pure state
 or an operator in entire space. Nevertheless, subregion complexity and 
 complexities for mixed states
 have also been studied in the context of holographic complexity 
 \cite{{Alishahiha:2015rta},{Carmi:2016wjl},{Agon:2018zso},
 {Alishahiha:2018lfv}}.   The circuit complexity for mixed states in 
 open systems has been also studied in
 \cite{Bhattacharyya:2020iic,Bhattacharyya:2021fii,Bhattacharya:2022gbz,Bhattacharya:2022wlp,
 Bhattacharyya:2022rhm,Bhattacharjee:2022lzy}.

Since the  approach  used  in the literature to study complexity, mainly 
relied on the state, its generalizations to mixed states or to subregions  
are not straightforward. On the other hand, in this paper, our construction 
for complexity  (of pure state) 
is given in terms of a particular trace over the density matrix. An 
advantage to define complexity in terms of
density matrix is that it may be extended for the cases where we are dealing
with  subregion or mixed states. In these cases one would  expect that
the definition would be the same  and we just need to use reduced density 
matrix.
 
Let us consider a quantum system whose Hilbert space can be decomposed into
two parts  ${\cal H}={\cal H}_A\otimes {\cal H}_B$. The dynamics of the 
system 
is given by a Hamiltonian which, in general, may not be decomposed into two
parts acting on ${\cal H}_A$ and ${\cal H}_B$ separately. Therefore, even if 
we start
with a reference state which is separable, as times goes the state spreading 
makes it very complicated.

Let us start with a reference state $|\psi\rangle$ and construct the 
orthonormal,
ordered basis using the Hamiltonian of the system. Then it is rather 
straightforward
to compute the reduced label operator by taking trace over subsystem $B$
$\ell_A={\rm Tr}_B(\ell)$. One may also compute reduced density matrix at 
given time $\rho_A(t)={\rm Tr}_B(\rho(t))$. Then the subregion Krylov 
complexity is defined by 
\be
{\cal C}_A(t)={\rm Tr}_A(\ell_A\rho_A(t))\,.
\ee
It is important to note that although the time evolution of the density 
matrix is simple and follows from the Schr\"odinger equation
\be
\rho(t)=e^{iHt}\rho(0)e^{-iHt},
\ee
for the case of reduced density matrix it is rather involved  even  for 
 an initial density matrix which is factorized $\rho(0)=\rho_A(0)\otimes
 \rho_B(0)$ \cite{Lidar:2019}. More precisely, using the fact that the
 density matrix is positive and normalized, one may write $\rho_B(0)=
 \sum_\mu \lambda_\mu |\mu\rangle\langle \mu|$ which yields
 \cite{Lidar:2019}
 \be\label{ReducedRho}
 \rho_A(t)=\sum_{\mu,\nu}K_{\mu\nu}(t)\,\rho_A(0)\,K^\dagger_{\mu\nu}(t)
 \ee
 where $K_{\mu\nu}(t)$ known as the Kraus operators  are
\be
K_{\mu\nu}(t)=\sqrt{\lambda_\nu} \langle\mu|e^{iHt}|\nu\rangle\,.
\ee
 
 In general it is not an easy task to compute complexity for reduced density 
 matrix,
though for a special cases where the dynamics of the two subsystem is 
separable, one can expect to make some progress. 

 More generally, motivated by the operator-state mapping  of the previous 
 section, let us consider the following time dependent density matrix
\be
\rho(t)=\sum_{a,b,a',b'}\varrho_{ab}(t)\varrho^*_{a'b'}(t)
\,|E_a,E_b\rangle\langle E_{b'},E_{a'}|\,.
\ee
whose initial density is found by setting $t=0$. Then the reduced density 
matrix at given time reads
\be
\rho_r(t)=\sum_{a,a',c}\varrho_{ac}(t)\varrho^*_{a'c}(t)
\,|E_a\rangle\langle E_{a'}|\,,
\ee
by which the complexity is 
\be
{\cal C}(t)={\rm Tr}(\ell_r \rho_r(t))=
\sum_{a,a',c}\varrho_{ac}(t)\varrho^*_{a'c}(t) A(E_a,E_{a'})\,.
\ee
For the case of $\varrho_{ab}(t)=e^{i(E_a\pm E_b)t}\varrho_{ab}$ one finds
\be\label{Sub-KC}
{\cal C}(t)=
\sum_{a,a'}e^{i(E_a-E_{a'})t}\rho(E_a,E_{a'}) A(E_a,E_{a'})\,.
\ee
which has essentially the same form as that we have found for pure state 
complexity. 
Here $\rho(E_a,E_{a,})=\sum_c\varrho_{ac}\varrho^*_{a'c}$. Note
that in this expression, the functions $A(E_a,E_{a'})$ are matrix elements 
of the reduced label operator.

For the maximally entangled case, given by the state
\be
|\psi\rangle=\sum_{a} \varrho_a|E_a,E_a\rangle,
\ee
$\varrho_a$ being a function of $E_a$ with a possible phase, one can assume 
to have a separable dynamics as follows
\be
|\psi(t)\rangle=f_1(H_1,t)f_2(H_2,t)\,|\psi\rangle,
\ee
where $f_i$'s are unitary transformations representing the time evolutions  
of each subsystem. Then the reduced density matrix turns out to be 
independent of time and is given by 
\be
\rho_1(t)={\rm Tr}_2(\rho(t))=\sum_{a}
|\varrho_a|^2|E_a\rangle\langle E_a|\,.
 \ee
 Clearly, even though the subsystem specified by the reduced density matrix 
 are complex, the complexity remains constant
 \be
 {\cal C}(t)=\sum_a \rho(E_a)  A(E_a,E_a),
 \ee
 where $\rho(E_a)=|\varrho_a|^2$. Therefore one may conclude that having non-
 zero entanglement entropy  is not 
 enough to crate  complexity under unitary separable time evolutions and a 
 direct interaction is needed. In other words to get complexity growth for 
 subsystem the dynamics of entanglement is matter (see \eqref{ReducedRho}).

This feature of complexity demonstrated above hints at the interesting fact 
that in order to have a growth of complexity, it is not sufficient to be 
content with the factorized structure of the density matrix, but rather it 
needs some interaction. This can alternatively be attributed to a 
fundamentally non-factorized structure of the Hilbert space as well. In the 
context of the entanglement structure of a generic quantum system, this 
connection was discussed in \cite{Nogueira:2021ngh, Banerjee:2022jnv}. 

This 
underlines the precise realization of the ER = EPR paradigm 
\cite{Maldacena:2013xja}. In the context of complexity, one can expect a 
very similar conclusion following the discussion above. This opens up the 
possibility of describing the growth of complexity in terms of connected 
saddles like replica wormholes which in turn could in principle provide the 
precise geometric interpretation of the modified replica trick for 
complexity advocated in \cite{Alishahiha:2022kzc, Alishahiha:2022exn}. Work 
in this direction is in progress and we hope to report on this soon.

As a final comment, we note that  the
holographic subregion complexity for a time-dependent geometry, given 
by the Vaidya metric, has been computed in \cite{Chen:2018mcc,Ling:2019ien,Auzzi:2019mah}. In the context of 
gauge-gravity duality, the Vaidya metric provides a holographic description
for a thermal quench. It was then possible to explore the time-dependence of holographic subregion complexity in this 
context. It was shown that the holographic subregion complexity 
exhibits linear growth up to a maximum value after which it shows a decreasing phase and finally saturates to a constant.

Intuitively, one can see that the Krylov subregion complexity 
should exhibit similar features. In fact, from equation \eqref{Sub-KC} one can note that the linear growth can indeed be understood from the pole structure of the 
$A$-function, much similar to what we studied in the context of Krylov complexity.
On the other hand, since the information about the Lanczos coefficients is encoded in the 
$A$-function, the saturation phase can be understood from the fact that the Lanczos coefficients vanish at late times.

It would, of course, be interesting to study Krylov subregion complexity more rigorously and compare it with that of holographic one explicitly.


\section{Conclusions}
\label{sec5}

To summarize our progress in this paper, we developed a unified formalism to 
study the state and the operator complexities. The key observation towards 
the unification is to realize that, within the realm of quantum mechanics, 
one can think of a channel-state correspondence which helps us to cast the 
the problem of studying operator complexity, to a corresponding state complexity obtained through the evolution of the dual state living on a doubled Hilbert space. 

While this connection itself is quite interesting and thought-provoking, it reveals, on its way to development, a couple of more surprises. First, it connects very naturally to the notion of complexity discussed in the context of holography following two routes - i) realizing the doubling of the 
Hilbert space for a more generic quantum system in terms of the entanglement structure arising from the Reeh-Schlieder theorem in an axiomatic quantum field theory, thus connecting this procedure to the state-dependent bulk reconstructions\cite{Papadodimas:2013jku}. This shows that the state-channel map which connects the state and operator complexities through a doubling of the Hilbert space, 
does secretly exploit the entanglement algebra imposed through the Lanczos algorithm. This hidden structure of entanglement underlying the Krylov construction, to our knowledge, had not been spelt out in the literature; ii) through a direct 
identification of the matrix elements of the label operator in the energy basis defined to quantify the spreading in a generic quantum system to the expectation values of the holographic position operator appearing in the quenched geodesic length as in \cite{Alishahiha:2022exn, Alishahiha:2022kzc, Alishahiha:2022nhe}. This provides a support for a general definition of complexity in terms of the pole structure of the $A$-function as advocated in \cite{Alishahiha:2022nhe}. 
Our construction establishes that this definition of complexity is universally applicable for any quantum system, with or without gravity, so long as the complexity has a late time linear growth. Such behaviour of complexity is indeed expected for a large class of physical systems, both integrable and chaotic and unlike the saturation phase, it is solely governed by the saturation of the Lanczos coefficients, captured by the pole structure of the $A$-function.

While these two routes lead to a further unification, now also incorporating the holographic complexity along with the state and the operator complexities, there is also a second but very important practical advancement in terms of developing a formalism to study subregion complexity and complexity for mixed quantum states, which had so far been quite illusive in the existing literature. In our formalism, this appears, rather naturally, since our universal definition of complexity is given in terms of a particular trace over the density matrix. In particular, the subregion complexity can be obtained simply 
by replacing the density matrix with the reduced 
density matrix corresponding to any given subregion. 
Furthermore, while generalizing the notion of complexity for a mixed state, we realise that this naturally hints at having a replica wormhole saddles in the dual gravitational spacetime, in line with the expectation coming from the modified replica trick proposed in \cite{Alishahiha:2022exn, 
Alishahiha:2022kzc}.

Following this short summary let us now conclude with a few ongoing progresses.

\paragraph{The mutual complexity} Since we are dealing with subregion 
complexity,
one can naturally define the Krylov mutual complexity for two subregions $A$ 
and $B$ as follows
\be
{\cal M}_{AB}={\cal C}_A+{\cal C}_B-{\cal C}_{A\cup B}\,,
\ee
where ${\cal C}$'s are the Krylov subregion complexities associated with the 
mentioned
regions. In the context of the holographic complexity, the mutual complexity 
has been  defined in \cite{Alishahiha:2018lfv,Caceres:2019pgf}.

\paragraph{Complexity in open quantum systems} Our universal formalism for 
state and operator complexities is applicable as well for open quantum 
systems which lack a Hermitian Hamiltonian which in turn means that the 
ordered basis defined in \eqref{GS} does not span the full doubled Hilbert 
space ${\cal H}_\psi$. In this case, one needs two independent sets of basis 
vectors, $|n\rangle$ and $\langle{\bar n}|$ which are not in general 
orthogonal to each other. Nevertheless, given a non-Hermitian operator $\cal 
A$, in many cases \footnote{Although this formalism was used and tested for 
pseudo-Hermitian Hamiltonians with real eigenvalues 
\cite{Mostafazadeh:2001jk, Gruning:2011}, we expect this procedure to work 
for more general non-Hermitian operators.}, it is possible to construct an 
anti-linear operator $\cal L$ such that ${\cal L}^{-1} {\cal A} \, {\cal L}$ 
is Hermitian with respect to the inner product $\langle n|
{\bar n}\rangle_{\cal L} \equiv \langle n|{\cal L} \,{\bar n}\rangle$. One 
example of such an anti-linear operator is the CPT operator. Adopting this 
formalism for computing complexity due to a non-Hermitian operator $\cal A$ 
amounts to a generalization of the definition of complexity \eqref{SpN} as

\be\label{SpN-mod}
{\cal C}_\phi={\rm Tr}(\ell \rho_\phi)_{\cal L}\,,
\ee
for any state $|\phi\rangle\in {\cal H}_\psi$.
Here the subscript $\cal L$ reminds us that the trace operation here should 
be performed taking into account the modified inner product $\langle n|
{\bar n}\rangle_{\cal L}$.

\paragraph{The transition matrix} As an aside comment, let us mention that 
there is an other operator, beside the
density matrix, which plays an important role in this context that is  known 
as the transition matrix 
\be
{\tau}=|\psi\rangle\langle \phi|\,.
\ee
It is natural to consider the case where the state $|\phi\rangle$ is given 
by \eqref{UA}. Although the information of Lanczos coefficients is encoded 
in the trace of the transition  matrix ${\rm Tr}(\tau)=\langle \psi|
U^\dagger({\cal A},s)|\psi\rangle
=\psi_0^*(s)$,
its evolution does not lead to any spreading as defined by 
\eqref{complexity}. More 
precisely one has
\be
{\rm Tr}(\ell \tau)=0.
\ee
It is important to note that the vanishing of the above equation is a direct
consequence of our definition of the label operator in which we set $c_n=n$ 
that is zero for $n=0$. If one assumes a general $c_n$ for $c_0\neq 0$ one 
has ${\rm Tr}(\ell \tau)=c_0 {\rm Tr}(\tau)$. 

Transition matrix is used to define the pseudo-entropy which is the 
generalization of entanglement entropy with post-selection 
\cite{Mollabashi:2020yie, Mollabashi:2021xsd}. 
 Taking into account that the fact that the autocorrelation can be
re-expressed in terms of the  the Lanczos coefficients, one would expect 
that  the pseudo-entropy exhibits certain universal behavior, at least for 
chaotic systems. We hope to report on this soon. 

\acknowledgements{We would like to thank Rathindra Nath Das, Moritz Dorband and Johanna Erdmenger for several useful and inspiring discussions.}

\bibliography{literature}

\end{document}